# ANHARMONIC DECAY OF PHONONS IN SEMICONDUCTORS FROM FIRST-PRINCIPLES CALCULATIONS

ALBERTO DEBERNARDI and STEFANO BARONI

Scuola Internazionale Superiore di Studi Avanzati (SISSA), Via Beirut 2-4, I-34014 Trieste, Italy

and

ELISA MOLINARI

Dipartimento di Fisica, Università di Modena, Via Campi 213/a, I-41100 Modena, Italy

The anharmonic contribution to phonon lifetimes and its temperature dependence is calculated from first principles in C, Si, and Ge using third-order density-functional perturbation theory. Good agreement with available experimental data is obtained. Different competing two-phonon decay channels are compared and correlated with the density of final states.

The anharmonic decay of phonons into modes of lower energy determines the homogeneous width of first-order Raman lines and influences a number of other important effects in semiconductors, including the decay rate of optical phonon populations as probed by time-resolved experiments.[1] Most of the theoretical studies performed so far in this field rely on simplifying assumptions on the possible decay mechanisms,[2] and/or on phenomenological models for describing harmonic and anharmonic interactions.[3] As discussed in Ref. 4, the ability of these schemes to account for experimental findings is however rather questionable. In recent years, first-principles techniques based on density-functional theory have shown to be a very accurate and computationally viable tool for predicting various vibrational properties of semiconductors. The use of these techniques to study the anharmonic decay of phonons would therefore be of great interest.[5] Here we tackle this problem using third-order density-functional perturbation theory.[6−9] Our results are in very good agreement with some of the available experimental data, which however show a considerable spread.

If only three-phonon processes are considered, energy and momentum conservation dictates that the LTO phonon decays into a pair of phonons with opposite momenta, whose frequencies sum up to the frequency of the decaying mode. The inverse lifetime of the LTO mode reads:[4]

$$\Gamma = \frac{18\pi}{\hbar^2} \sum_{\mathbf{q},j_1,j_2} \left| V(\mathbf{0}j, \mathbf{q}j_1, -\mathbf{q}j_2) \right|^2 \left( n_{j_1}(\mathbf{q}) + n_{j_2}(-\mathbf{q}) + 1 \right) \times$$
$$\delta\left( \omega_{LTO} - \omega_{j_1}(\mathbf{q}) - \omega_{j_2}(-\mathbf{q}) \right), \quad (1)$$

where $n$ are the thermal occupation numbers, the $j$'s indicate the phonon branches ($j = 1 \div 6$), and $V$ are anharmonic coupling coefficients, as defined e.g. in Ref. 4.

Phonon inverse lifetimes, Eq. (1), have been obtained using phonon dispersions from Refs. 7 and 10, and anharmonic coupling coefficients calculated following Ref. 9. The calculation of anharmonic couplings makes use of the so called $2n + 1$ *theorem* which states that the knowledge of the electronic charge-density response of a system up to order $n$ in the strength of an external perturbation is sufficient to determine the energy derivative with respect to the strength up to order $2n + 1$.[11] For $n = 1$, this theorem ensures that third-order anharmonic couplings can be calculated from the *linear* response of the electron density distribution to lattice distortions, hence from the same ingredients which enter standard lattice-dynamical calculations in the harmonic approximation.[7] Our calculations have been performed within density-functional theory in the local-density approximation and using the plane-wave pseudopotential method. All the technical details are the same as in Refs. 7 and 10.

**Table 1.** Calculated full widths at half maximum ($2\Gamma$) of zone-center optical phonons. The corresponding low-temperature experimental values are shown for comparison. The last columns indicate the relative contribution to $\Gamma$ of the different decay channels.

|    | $2\Gamma$ (cm$^{-1}$) | $2\Gamma$ (expt) (cm$^{-1}$) | LA+LA (%) | LA+TA (%) | TA+TA (%) |
|----|-----------------------|------------------------------|-----------|-----------|-----------|
| C  | 1.29 | 1.2[a]  | 15.4 | 34.6 | 50.0 |
| Si | 1.43 | 1.45[b] | 5.7  | 94.3 | —    |
| Ge | 0.73 | 0.75[c] | 4.3  | 95.7 | —    |

[a] Ref. 12*(f)*; [b] Ref. 13*(c)*; [c] Ref. 4.

In Table 1 we report our calculated linewidths of the LTO phonons in C, Si, and Ge. The spread in the corresponding experimental data is rather large, ranging from 1.2 to 2.9 cm$^{-1}$ for C,[12] from 1.24 to 2.1 cm$^{-1}$ for Si,[13] and from 0.75 to 1.4 cm$^{-1}$ for Ge.[14] Among these, in Table 1 we display the values which are closer to our calculations, resulting in a very good agreement.

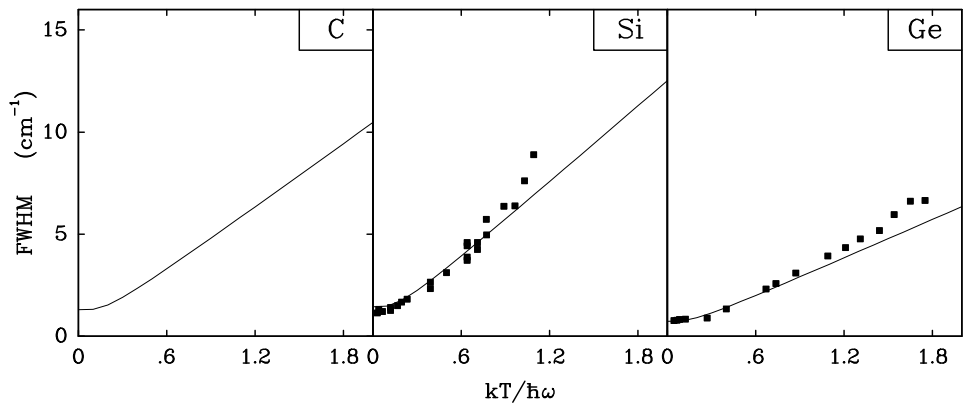

**Figure 1.** Temperature dependence of the full width at half maximum of the LTO phonon line in C, Si and Ge. Solid lines are the result of the present calculation; squares represent experimental data from Ref. 4.

The temperature dependence of the phonon linewidth is given by Eq. (1) through the occupation numbers. The results are presented in Fig. 1 together with the available experimental data. The agreement is very good up to temperatures $T \gtrsim \hbar \omega_{LTO}/k_B$. Above these temperatures higher-order anharmonic terms are likely to account for the discrepancies.

By restricting the sums over the $j$'s to selected final states in Eq. (1), one can identify which are the relevant decay processes. The results of this analysis are also reported in Table 1, where the decay channels are indicated by 'TA' ($j = 1, 2$) and 'LA' ($j = 3$). The decay into one optical and one acoustic phonon is kinematically forbidden in all the three cases examined. For Si and Ge, the dominant channel involves one LA and and one TA mode as final states, consistently with the results of Ref. 5. The 'Klemens channel', i.e. the decay of the LTO mode into two acoustic phonons belonging to a same branch and with opposite momenta, turns out to give a very small contribution in these materials. The situation is quite different in diamond where the TA+TA channel becomes dominant at the expenses of the LA+TA channel. In order to obtain more detailed information on the relevant decay mechanisms, we decompose Eq. (1) into contributions, $\gamma(\omega)$, from different regions of the frequency spectrum. The function $\gamma(\omega)$ is defined by an equation similar to Eq. (1) where the sum over $j_1$ and $\mathbf{q}$ is restricted to those values for which $\omega_{j_1}(\mathbf{q}) = \omega$. This amounts to inserting $\delta(\omega - \omega_{j_1}(\mathbf{q}))$ under the sign of sum. Note that $\int \gamma(\omega) d\omega = \Gamma$.

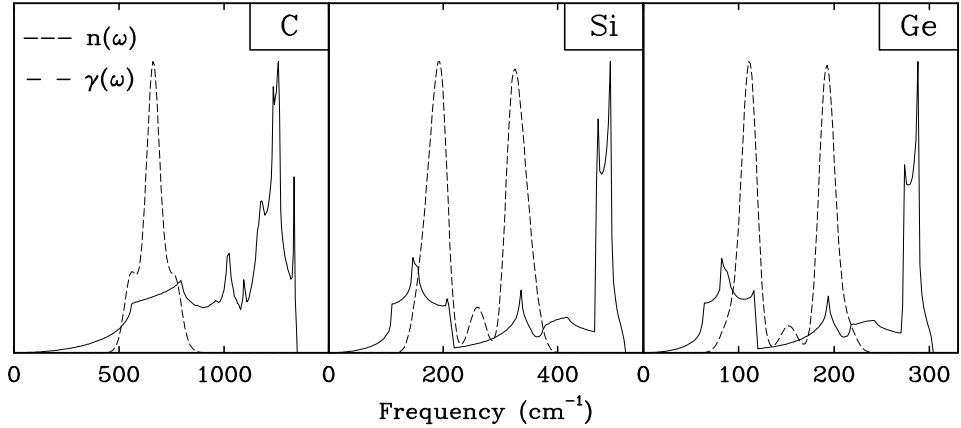

**Figure 2.** Calculated phonon density of states, $n(\omega)$ (solid line), and spectral decomposition of the phonon linewidths, $\gamma(\omega)$ (dashed line), for the three elemental semiconductors C, Si and Ge.

In Fig. 2 we display $\gamma(\omega)$ and compare it with the one-phonon density of states (DOS), $n(\omega)$.[15] By definition, $\gamma(\omega)$ is symmetric around $\omega_{LTO}/2$. At this frequency, $\gamma(\omega)$ displays a peak corresponding to the Klemens decay mechanism. Lateral peaks may occur symmetrically with respect to $\omega_{LTO}/2$ at frequencies $\omega_1$ and $\omega_2$ such that $n(\omega_1)$ and $n(\omega_2)$ are large and arise from a *same* region of the Brillouin zone. In diamond the Klemens peak is dominant because $\omega_{LTO}/2$ falls in a region where the DOS is relatively large (between $\omega_{TA(L)}$ and $\omega_{TA(X)}$). The lateral peaks are less intense because the above conditions are not fulfilled in diamond. The opposite is true for Si and Ge, where $n(\omega_{LTO}/2)$ is very small (and so is therefore the central peak). A careful analysis of the phonon dispersion relations reveals that the main contribution to the lateral peaks comes from the decay into modes with $j = 2$ and $j = 3$ around the WL symmetry line, in agreement with the suggestions of Ref. 4.

Our results show that third-order interactions provide an accurate description of the anharmonic decay of phonons in semiconductors up to far above room temperature. These interactions can be conveniently obtained from first principles by density functional perturbation theory with an effort comparable to that required for standard lattice-dynamical calculations in the harmonic approximations. This opens the way to the prediction of anharmonic lifetimes in systems—such as some bulk compound semiconductors (e.g. AlAs) or heterostructures—where they are not easily accessed by experiments.

**References**

1. For a recent review see: J.A. Kash and J.C. Tsang, in *Light Scattering in Solids VI*, edited by M. Cardona and G. Güntherodt (Springer, Berlin, 1991), p. 423.
2. P.G. Klemens, Phys. Rev. **148**, 845 (1966).
3. R.A. Cowley, J. Phys. (Paris) **26**, 659 (1965).
4. J. Menéndez and M. Cardona, Phys. Rev. B **29**, 2051 (1984).
5. The only previous attempt to apply ab-initio techniques to the anharmonic decay of phonons was done for Si using a semiempirical lattice-dynamical model fitted to a few frozen-phonon calculations: S. Narasimhan and D. Vanderbilt, Phys. Rev. B **43**, 4541 (1991).
6. S. Baroni, P. Giannozzi, and A. Testa, Phys. Rev. Lett. **58**, 1861 (1987).
7. P. Giannozzi, S. de Gironcoli, P. Pavone, and S. Baroni, Phys. Rev. B **43**, 7231 (1991).
8. X. Gonze and J.-P. Vigneron, Phys. Rev. B **39**, 13120 (1989).
9. A. Debernardi and S. Baroni, Solid State Commun. **91**, 813(1994).
10. P. Pavone, K. Karch, O. Schütt, D. Strauch, W. Windl, P. Giannozzi, and S. Baroni, Phys. Rev. B **48**, 3156 (1993).
11. The $2n+1$ theorem is known since long time in elementary quantum mechanics. See e.g. P. M. Morse and H. Feshbach, *Methods of Theoretical Physics* (Mc Graw-Hill, New York, 1953), Vol. II, P. 1120. Recently, the validity of this theorem has been extended to density-functional theory in Ref. 8.
12. *(a)* R.S. Krishnan, Proc. Indian Acad. Sci. **24**, 45 (1946); *(b)* S.A. Solin and A.K. Ramdas, Phys. Rev. B **1**, 1687 (1970); *(c)* E. Anastassakis, H.C. Hwang, and C.H. Perry, Phys. Rev. B **4**, 2493 (1971); *(d)* W.J. Borer, S.S. Mitra, and K.W. Namjoshi, Solid State Commun. **9**, 1377 (1971); *(e)* R.M. Chrenko, J. Appl. Phys. **63**, 5873 (1988); *(f)* M.A. Washington and H.Z. Cummins, Phys. Rev. B **15**, 5840 (1977); *(g)* K.C. Hass, M.A. Tamor, T.R. Anthony, and W.F. Banholzer, Phys. Rev. B **44**, 12046 (1991); *(h)* J. Spitzer, P. Etchegoin, M. Cardona, T.R. Anthony, and W.F. Banholzer, Solid State Commun. **88**, 509 (1993).
13. *(a)* T.R. Hart, R.L. Aggarwal, and B. Lax, Phys. Rev. B **1**, 638 (1970); *(b)* R. Tubino, L. Piseri, and G. Zerbi, J. Chem. Phys. **56**, 1022 (1972); *(c)* P.A. Temple and C.E. Hathaway, Phys. Rev. B **7**, 3685 (1973); *(d)* M. Balkanski, R.F. Wallis, and E. Haro, Phys. Rev. B **28**, 1928 (1983); *(e)* Ref. 4.
14. *(a)* R.K. Ray, R.L. Aggarwal, and B. Lax, in *Light Scattering in Solids*, edited by M. Balkanski (Flammarion, Paris, 1971), p. 288; *(b)* F. Cerdeira and M. Cardona, Phys. Rev. B **5**, 1440 (1972); *(c)* Ref. 4; *(d)* H.D. Fuchs, C.H. Grein, R.I. Devlen, J. Kuhl, and M. Cardona, Phys. Rev. B **44**, 8633 (1991).
15. The actual calculation of $\gamma(\omega)$ has been performed by broadening $\delta(\omega-\omega_j)$ with a gaussian whose variance is a few cm$^{-1}$.